\documentclass[aip,rsi,amsmath,amssymb,reprint]{revtex4-1}

\usepackage{graphicx}
\usepackage{dcolumn}
\usepackage{bm}
\usepackage{csquotes}
\usepackage{epstopdf}
\usepackage{multirow}
\usepackage[breaklinks=true]{hyperref}
\usepackage{breakurl}
\hypersetup{colorlinks, linkcolor=red, citecolor=red, urlcolor=blue}

\begin{document}

\preprint{AIP/123-QED}

\title[Huddle test measurement of a near Johnson noise limited geophone]{Huddle test measurement of a near Johnson noise limited geophone}

\author{R. Kirchhoff}
 \email{robin.kirchhoff@aei.mpg.de}
 \affiliation{Albert-Einstein-Institute / Max-Planck-Institute for Gravitational Physics,\\ 
 D-30167 Hanover, Germany}
 \affiliation{Leibniz Universit\"at Hannover, D-30167 Hanover, Germany}
\author{C. M. Mow-Lowry}
\affiliation{School of Physics and Astronomy and Institute of Gravitational Wave Astronomy, University of Birmingham, Edgbaston, Birmingham B15 2TT, United Kingdom}
\author{V. B. Adya}
 \affiliation{Albert-Einstein-Institute / Max-Planck-Institute for Gravitational Physics,\\ 
 D-30167 Hanover, Germany}
\affiliation{Leibniz Universit\"at Hannover, D-30167 Hanover, Germany}
\author{G. Bergmann}
\affiliation{Albert-Einstein-Institute / Max-Planck-Institute for Gravitational Physics,\\ 
 D-30167 Hanover, Germany}
 \affiliation{Leibniz Universit\"at Hannover, D-30167 Hanover, Germany}
\author{S. Cooper}
\affiliation{School of Physics and Astronomy and Institute of Gravitational Wave Astronomy, University of Birmingham, Edgbaston, Birmingham B15 2TT, United Kingdom}
\author{M. M. Hanke}
\affiliation{Albert-Einstein-Institute / Max-Planck-Institute for Gravitational Physics,\\ 
 D-30167 Hanover, Germany}
 \affiliation{Leibniz Universit\"at Hannover, D-30167 Hanover, Germany}
\author{P. Koch}
\affiliation{Albert-Einstein-Institute / Max-Planck-Institute for Gravitational Physics,\\ 
 D-30167 Hanover, Germany}
 \affiliation{Leibniz Universit\"at Hannover, D-30167 Hanover, Germany}
\author{S. M. K\"ohlenbeck}
\affiliation{Albert-Einstein-Institute / Max-Planck-Institute for Gravitational Physics,\\ 
 D-30167 Hanover, Germany}
 \affiliation{Leibniz Universit\"at Hannover, D-30167 Hanover, Germany}
\author{J. Lehmann}
\affiliation{Albert-Einstein-Institute / Max-Planck-Institute for Gravitational Physics,\\ 
 D-30167 Hanover, Germany}
 \affiliation{Leibniz Universit\"at Hannover, D-30167 Hanover, Germany}
 \author{P. Oppermann}
 \affiliation{Albert-Einstein-Institute / Max-Planck-Institute for Gravitational Physics,\\ 
 D-30167 Hanover, Germany}
 \affiliation{Leibniz Universit\"at Hannover, D-30167 Hanover, Germany}
\author{J. W\"ohler}
\affiliation{Albert-Einstein-Institute / Max-Planck-Institute for Gravitational Physics,\\ 
 D-30167 Hanover, Germany}
 \affiliation{Leibniz Universit\"at Hannover, D-30167 Hanover, Germany}
\author{D. S. Wu}
\affiliation{Albert-Einstein-Institute / Max-Planck-Institute for Gravitational Physics,\\ 
 D-30167 Hanover, Germany}
 \affiliation{Leibniz Universit\"at Hannover, D-30167 Hanover, Germany}
\author{H. L\"uck}
\affiliation{Albert-Einstein-Institute / Max-Planck-Institute for Gravitational Physics,\\ 
 D-30167 Hanover, Germany}
\affiliation{Leibniz Universit\"at Hannover, D-30167 Hanover, Germany}
\author{K. A. Strain}
\affiliation{Albert-Einstein-Institute / Max-Planck-Institute for Gravitational Physics,\\ 
 D-30167 Hanover, Germany}
\affiliation{Scottish Universities Physics Alliance, School of Physics and Astronomy, University of Glasgow, Glasgow G12 8QQ, United Kingdom}

\date{\today}

\begin{abstract} 
In this paper the sensor noise of two geophone configurations (L-22D and L-4C geophones from Sercel with custom built amplifiers) was measured by performing two huddle tests. It is shown that the accuracy of the results can be significantly improved by performing the huddle test in a seismically quiet environment and by using a large number of reference sensors to remove the seismic foreground signal from the data. Using these two techniques, the measured sensor noise of the two geophone configurations matched calculated predictions remarkably well in the bandwidth of interest (0.01\,Hz to 100\,Hz). Low noise operational amplifiers OPA188 were utilized to amplify the L-4C geophone to give a sensor that was characterized to be near Johnson noise limited in the bandwidth of interest with a noise value of $10^{-11}\,\text{m}/\sqrt{\text{Hz}}$ at 1\,Hz.
\end{abstract}

\maketitle

\section{\label{Introduction}Introduction}

Measurements of seismic motion are an important technique in various branches of modern science and engineering. In particular, low frequency measurements ($\lesssim$ 1\,Hz) of motion are essential for studying natural phenomena such as earthquakes \cite{Arif2011}, the seafloor \cite{Wayne2000}, improved searches for minerals, oil and gas\cite{Eaton2003,Saenger2009} and the advanced prediction of the stability of underground structures \cite{Hashasha2001} or other buildings \cite{Huston2010}. In addition, these measurements are also crucial for the seismic isolation of large scale, high precision instruments such as gravitational wave interferometers (LIGO \cite{Matichard2015}, VIRGO \cite{Acernese2015}, KAGRA \cite{Fujii2016}) and particle colliders (CLIC \cite{Balik2014}).\\\\
The Albert-Einstein-Institute 10\,m Prototype Facility develops and tests novel techniques for the gravitational wave interferometry\cite{Westphal2012}. The isolation of optics from ground motion is crucial for the sensitivity of gravitational wave detectors. In the 10\,m Prototype Facility we use seismic attenuation systems (AEI-SASs) that combine passive and active isolation techniques to decouple optical tables from seismic motion. Detailed explanations of the AEI-SAS can be found in [Wanner \textit{et al.}\cite{Wanner2013,Wanner2012}] and [Bergmann \textit{et al.}\cite{Bergmann2017}]. To implement active control, a multitude of sensors is needed to measure the motion on the optical table in every degree of freedom. Reviews of seismic sensors can be found in [Collette \textit{et al}.\cite{Collette2012}] and [Wielandt\cite{Wielandt2002}], including descriptions of the geophones, accelerometers and seismometers which are utilized in the AEI-SAS. A description of the usage and the positioning of the sensors in the AEI-SAS can be found in [Wanner \textit{et al.}\cite{Wanner2013,Wanner2012}]. Geophones are used to measure inertial vertical motion and tilt. The noise of the geophones and their amplifier electronics will ultimately limit the performance of a control loop motivating us to investigate the noise performance of geophones and their electronics.\\\\
The principle operation of a geophone is based on a harmonic oscillator granting sensitivity for inertial motion in a single direction. The readout is based on an inductive generation of a current inside a coil moving relative to a magnet. The signal of a geophone $x_{s}(t)$ is determined by the sum of the detected motion $x_{m}(t)$ acting on this sensor and the sensor noise $x_{n}(t)$ as
\begin{eqnarray}
x_{s}(t) = x_{m}(t) + x_{n}(t).
\label{3.1}
\end{eqnarray}
At frequencies below $\approx$\,0.1\,Hz to 5\,Hz (the exact frequency depends on the noise level of the geophone and its amplifier electronics), the geophone output is dominated by its sensor noise. At higher frequencies large foreground signals resulting from the motion of the geophone dominate the sensor noise making the geophone a good seismic sensor.\\\\
As written above it can also be interesting to measure the noise of a geophone and its amplifier electronics. This paper will demonstrate a huddle test, which is a technique to remove the large foreground signal and to reveal the underlying sensor noise. Multiple sensors are utilized to enable coherent subtraction of the common foreground signal. This measurement technique is used to compare the sensor noise of the L-22D and the L-4C geophones (manufactured by Sercel \cite{Sercel2016}) and their amplifiers. The L-4C geophone and amplifier configuration presented here is shown to be near Johnson noise limited in the range of 0.01\,Hz and 100\,Hz. 

\section{Geophone noise sources}
\label{Geophone noise sources}

\begin{figure}
\centering
\includegraphics[trim = 0cm 0cm 0cm 0cm, clip, width=8.5cm]{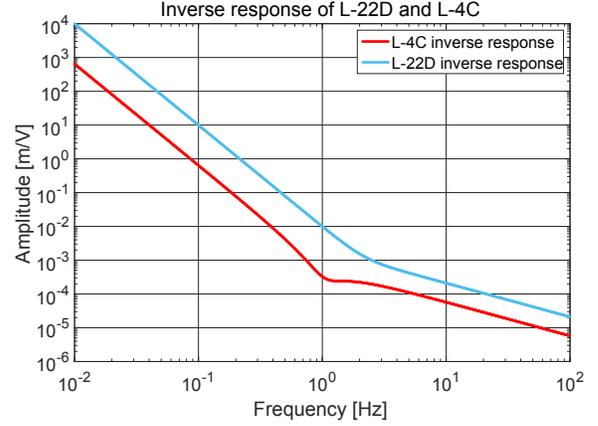}
\caption{\footnotesize{The inverse response functions of the L-22D and the L-4C geophone and their amplifier electronics. As the L-4C has a lower resonance frequency and a higher sensitivity, its inverse response is lower over the whole bandwidth of interest (0.01\,Hz to 100\,Hz).}}
\label{inverseresponse}
\end{figure}

The output signal directly from a geophone is often too small to be measured using data acquisition devices and a low-noise preamplifier is typically employed. This has the additional benefit of buffering the geophone i.e. making the operation of the geophone independent from the input impedance of the connected recording system. Two geophone amplifier circuits were constructed and employed in these tests. Their schematics are shown in the supplementary materials (section \ref{Supplementary material}). A calculation of the total noise of a geophone and its amplifier electronics is demonstrated in [Barzilai \textit{et al.}\cite{Barzilai1998}] by calculating the uncorrelated sum of the various individual contributions. Here, slightly modified equations are used to calculate the input referred noise as displacement equivalent noise spectral densities ($\tilde{n}$) in units of $\text{m}/\sqrt{\text{Hz}}$. This is achieved by multiplying the measured noise with the inverse response (FIG. \ref{inverseresponse}) of the geophone and its amplifier electronics. The response is defined as the transfer function from the motion acting on the geophone to the signal that is measured. It consists of the transfer function of the suspended mass inside the geophone multiplied by the geophone sensitivity and the gain of the amplifier circuit. Note that although geophones measure velocity, displacement is used here due to the intended application for gravitational wave detectors. \\\\
The total noise $\tilde{n}_{\text{total}}$ is approximately given by the incoherent sum of the suspension thermal noise of the geophone oscillator $\tilde{n}_{\text{s}}$, the Johnson noise of the geophone coil $\tilde{n}_{\text{J}}$, the voltage noise of the operational amplifier being used in the first gain stage of the amplifier electronics $\tilde{n}_{\text{v}}$ and the current noise of this amplifier $\tilde{n}_{\text{c}}$. For an angular frequency, $\omega$, this is expressed as
\begin{eqnarray}
\tilde{n}_{\text{total}}(\omega)^2 &= \tilde{n}_{\text{s}}(\omega)^2 + \tilde{n}_{\text{J}}(\omega)^2 + \tilde{n}_{\text{v}}(\omega)^2 + \tilde{n}_{\text{c}}(\omega)^2.
\label{3.2}
\end{eqnarray}
The individual terms are calculated by the following equations:\\
\begin{eqnarray}
\tilde{n}_{\text{s}}(\omega) &&= \sqrt{\dfrac{4k_{\text{b}}T\omega_0}{mQ}} \times \dfrac{4\pi^2}{\omega^2},\nonumber\\
\nonumber\\
\tilde{n}_{\text{J}}(\omega) &&= \sqrt{4k_{\text{b}} T \Re(Z) } \times \dfrac{1}{Resp(\omega)},\nonumber\\
\nonumber\\
\tilde{n}_{\text{v}}(\omega) &&= \dfrac{N_{\text{V}}(\omega)}{Resp(\omega)},\nonumber\\
\nonumber\\ 
\tilde{n}_{\text{c}}(\omega) &&= \dfrac{N_{\text{A}}(\omega) \Re(Z)}{Resp(\omega)}.
\label{3.3}
\end{eqnarray}
\begin{table*}
\renewcommand{\arraystretch}{1.5}
\begin{tabular}{|c||l|c|c|}
\hline
\textbf{Parameter} & \textbf{description} & \textbf{L-22D} & \textbf{L-4C}\\
\hline \hline
$T$ & temperature & \multicolumn{2}{c|}{300\,K} \\
\hline
$\omega_0$ & resonance frequency of the geophone oscillator & 2\,Hz & 1\,Hz \\
\hline
$m$ & suspended mass of the geophone &  0.0728\,Kg & 0.96\,Kg\\
\hline
$Q$ & quality factor of the geophone oscillator & 0.5 & 3 \\
\hline
$Resp(\omega)$ & response of the geophone & \multicolumn{2}{c|}{FIG. \ref{inverseresponse}}\\
\hline
$\Re(Z)$ & real part of the geophone impedance &\multicolumn{2}{c|}{FIG. \ref{impedance}} \\
\hline
$N_{\text{V}}(\omega)$ & input-referred voltage noise& \multirow{2}{*}{\centering INA128\cite{INA128}} & \multirow{2}{*}{\centering OPA188\cite{OPA188,Hoyland2016}}\\
\cline{1-2} 
$N_{\text{A}}(\omega)$ & input-referred current noise&  &  \\
\hline
\end{tabular}
\caption{The characteristics of the L-22D geophones amplified with the INA128 and the L-4C geophones amplified with the OPA188. The values were either measured or taken from the data sheets. $N_{\text{V}}$ and $N_{\text{A}}$ refer to the INA128 amplifier for the L-22D and to the OPA188 amplifier for the L-4C.}
\label{ta.1}
\end{table*}
\begin{figure}
\centering
\includegraphics[trim = 0cm 0cm 0cm 0cm, clip, width=8.5cm]{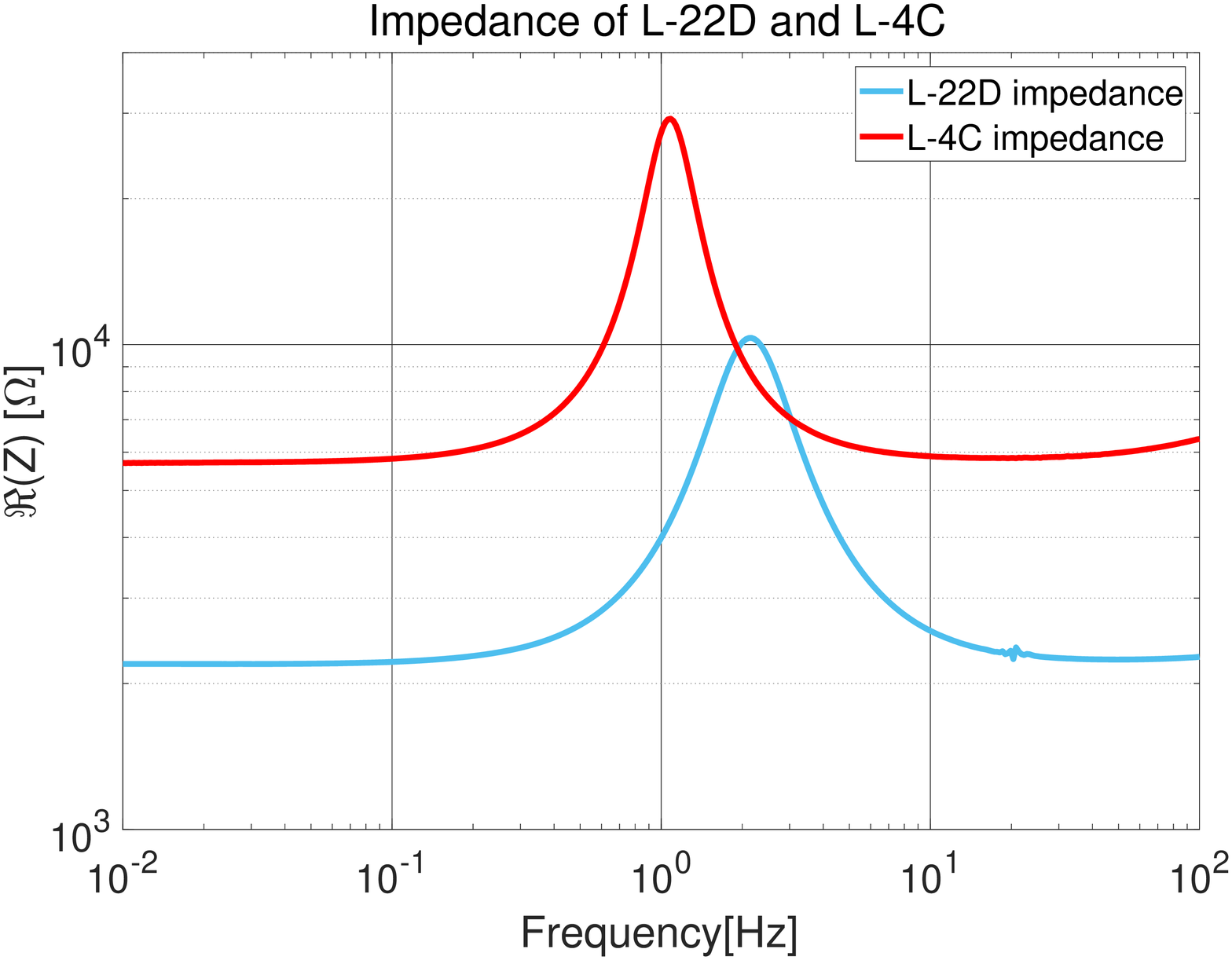}
\caption{\footnotesize{Real part of the impedance and the phase of the L-22D and the L-4C geophone. For this measurement a small voltage was sent through the geophones while they were exposed to ground motion.}}
\label{impedance}
\end{figure}
\begin{figure*}[hbtp]
\begin{minipage}[t]{8.5cm}
\centering
\includegraphics[trim = 0cm 0cm 0cm 0cm, clip, width=8.5cm]{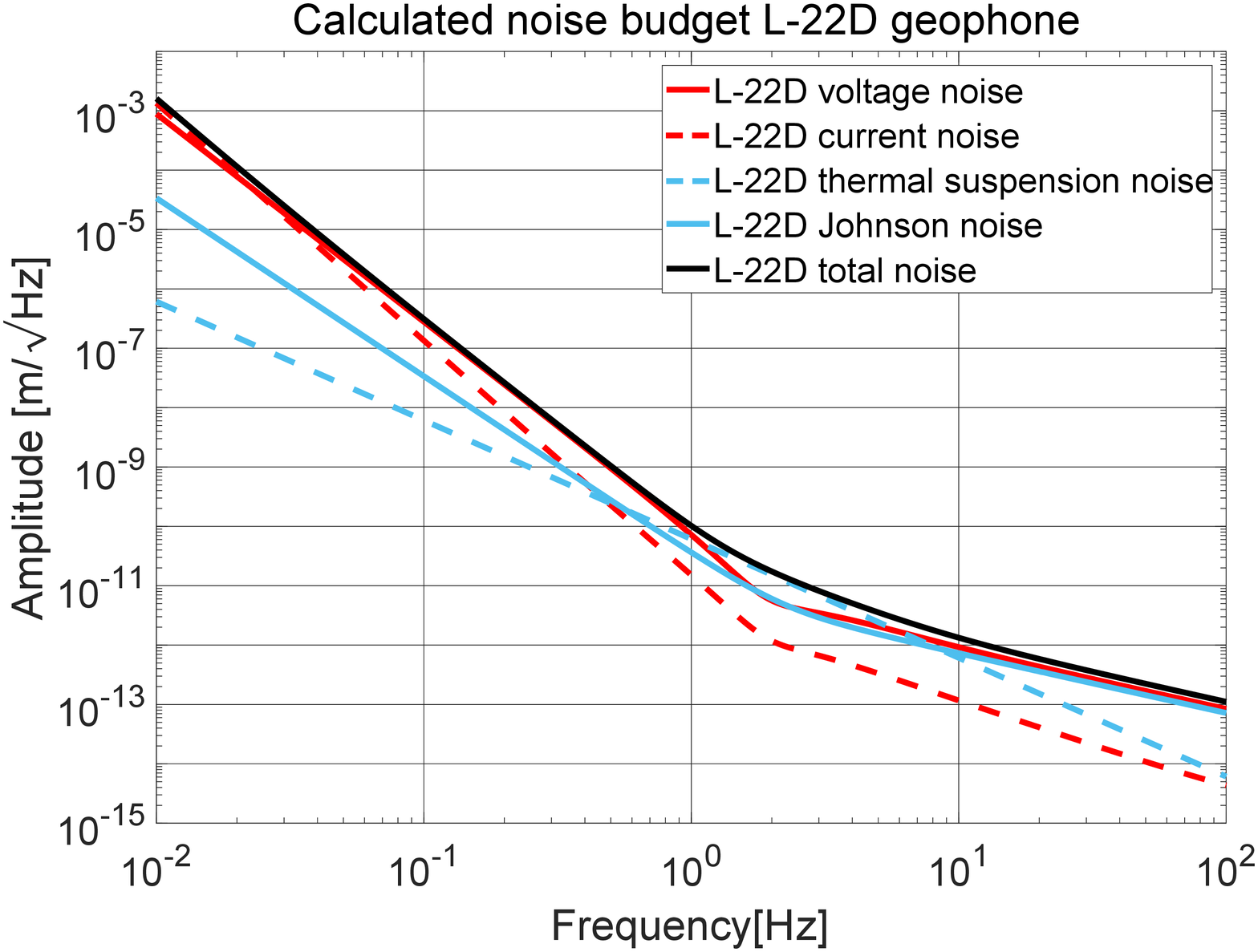}
\end{minipage}
\begin{minipage}[t]{8.5cm}
\centering
\includegraphics[trim = 0cm 0cm 0cm 0cm, clip, width=8.5cm]{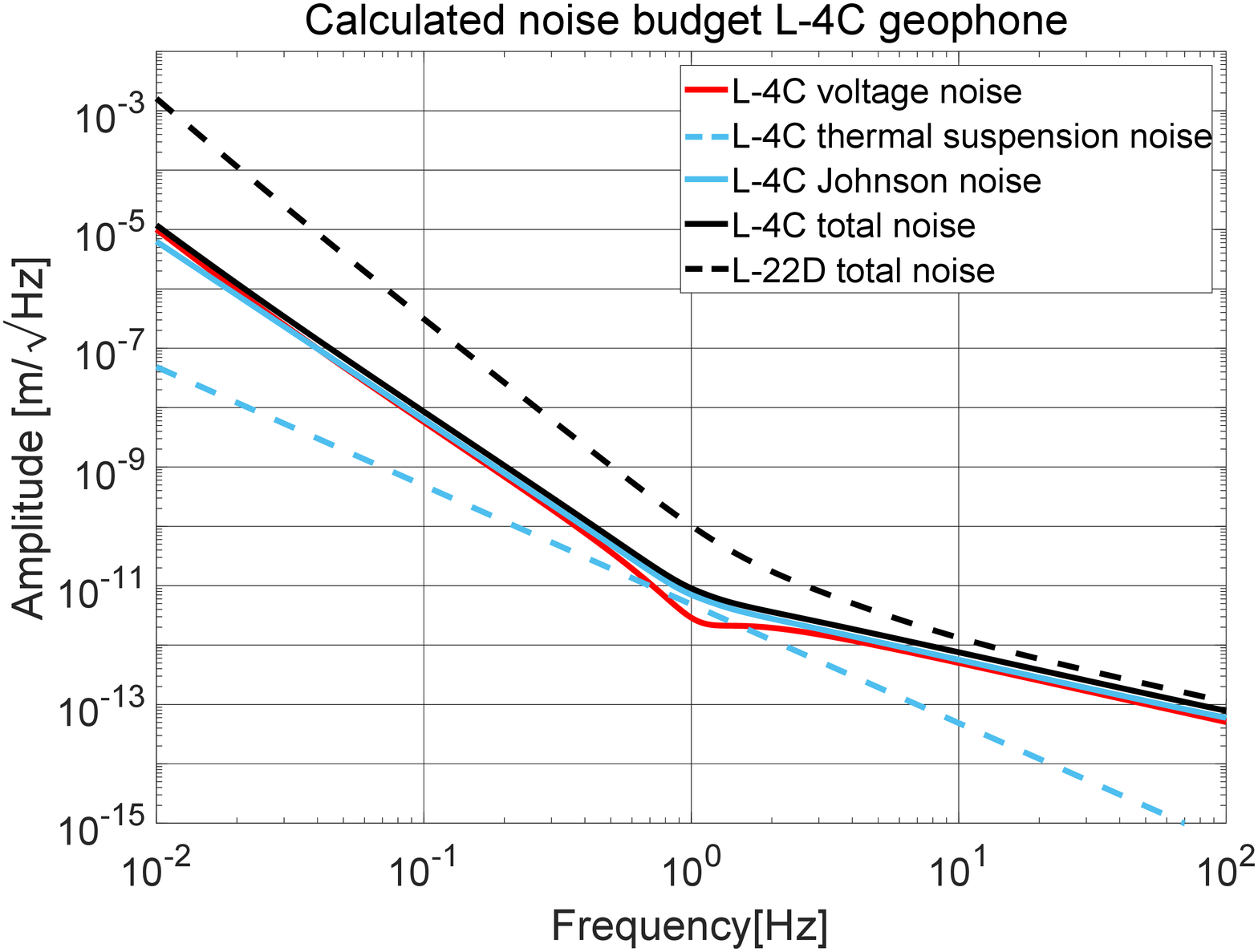}
\end{minipage}
\hfill
\begin{minipage}[t]{8.5cm}
\centering
\caption{The $\tilde{n}$ of the L-22D geophone with an INA128 first-stage amplifier. The total noise is dominated by current noise ($\lesssim$\,0.03\,Hz) and voltage noise (0.03\,Hz to 1\,Hz) at low frequencies, by suspension noise in the mid frequency regime (1\,Hz to 10\,Hz) and by Johnson noise of the geophone coil and voltage noise at high frequencies ($\gtrsim$\,10\,Hz). By exchanging the INA128 with the low noise amplifier OPA188 a significant improvement at low frequencies (factor of $\approx$\,10 at 0.1\,Hz) can be achieved.}
\label{L22old}
\end{minipage}
\hfill
\begin{minipage}[t]{8.5cm}
\centering
\caption{The $\tilde{n}$ of the L-4C geophone with an OPA188 first-stage amplifier. The total noise is dominated by voltage noise of the first stage amplifier and approximately equally by Johnson noise of the geophone coil over the whole frequency range. The total noise of the L-22D with the INA128 is shown for comparison.}
\label{L4Cnew}
\end{minipage}
\end{figure*} 
A detailed list of the above parameters for both geophones is given in TABLE \ref{ta.1}. In our configurations, the total Johnson noise is dominated by the geophone coil which allows us to neglect the contribution of other electronic components e.g.~feedback resistors. A measurement of the impedance of the L-22D and the L-4C geophone is shown in FIG. \ref{impedance}. A theoretical derivation is given in [Lantz\cite{Lantz2005}]. The voltage and current noise are assumed to be dominated by the first stage amplifier since it has a gain of 100. The key difference between the two amplifier designs is that the L-22Ds are amplified with INA128 whereas the L-4Cs are amplified with the OPA188 operational amplifiers. Their voltage and current noise can be found in their data sheets\cite{INA128,OPA188}. Additionally, direct noise measurements of the OPA188 can be found in [Hoyland\cite{Hoyland2016}]. The current noise of this amplifier could not be measured by [Hoyland\cite{Hoyland2016}]. However the values given in the data sheet \cite{OPA188} are low enough to be considered negligible for our purpose.\\\\
FIG. \ref{L22old} and FIG. \ref{L4Cnew} show the calculated noise budget for the L-22D and L-4C geophones and their amplifier electronics respectively. $\tilde{n}_{\text{total}}$ of the L-22D configuration is dominated by Johnson and voltage noise at high frequencies ($\gtrsim$\,10\,Hz), suspension noise in the mid frequency regime (1\,Hz to 10\,Hz), whereas voltage noise (0.03\,Hz to 1\,Hz) and current noise ($\lesssim$\,0.03\,Hz) dominate at low frequencies. The L-4C geophones themselves have lower noise than the L-22Ds across the entire frequency range due to a lower resonance frequency $\omega_0$, a stronger response $Resp(\omega)$, a larger mass $m$ and a higher quality factor $Q$. However, it can be seen that the L-4Cs amplified with the L-22D electronics would still be limited by amplifier noise at low frequencies. To achieve a significant improvement at frequencies below $\approx$\,1\,Hz, a new amplifier circuit was designed that has a low noise especially at low frequencies by using OPA188 operational amplifiers. From FIG. \ref{L4Cnew}, we expect $\tilde{n}_{\text{total}}$ of the L-4C configuration to be lower than $\tilde{n}_{\text{total}}$ of the L-22D configuration by a factor of 20 at 1\,Hz and by a factor of 40 at 0.1\,Hz. $\tilde{n}_{\text{total}}$ of the L-4C configuration is equally dominated by Johnson and voltage noise across our bandwidth of interest (0.01\,Hz to 100\,Hz).

\section{Huddle Test}
\label{Huddle Test}
As described in section \ref{Introduction}, it is difficult to precisely measure the geophone noise at low frequencies due to foreground seismic and anthropological signals which are significantly stronger than the noise. In order to evaluate the noise performance of a seismic sensor, it is common to perform a huddle test using multiple additional sensors located as close as possible to the target sensor to measure the common seismic motion. These additional sensors are referred to as reference sensors. The common signal will be coherent between all sensors, in contrast to the sensor noise which is incoherent. The data is processed to subtract the coherent common mode signal from the device under test. The remaining incoherent signal is assumed to be the sensor noise.\\\\
For the huddle test performed here, the recorded data was processed using a MATLAB script (Multi-Channel Coherent Subtraction, see supplementary material, section \ref{Supplementary material}) developed by Brian Lantz and Wensheng Hua based on the method presented by [Allen \textit{et al.}\cite{Allen1999}]. The software converts time series data from the multiple sensors into the frequency domain using Fast Fourier Transforms (FFT).\\\\
Recalling equation \ref{3.1}, a geophone signal is composed of the actual measured motion and the noise, which are inseparable. The actual motion can be calculated from the other sensors in the huddle test and subtracted from the total signal of the target sensor as
\begin{eqnarray}
\bar{X}_{\text{n}} (\omega) = X_{\text{s}} (\omega) -  \sum_{j=1}^{N} T_{j} (\omega) \times Y_{j}(\omega).
\label{4.2}
\end{eqnarray}
$\bar{X}_{\text{n}}$ and $X_{\text{s}}$ are the estimated noise and total signal from the geophone under test. $Y_{j}$ is the signal from the $j$-th reference sensor (out of a total $N$). $T_{j}$ is the transfer function which accounts for the differences in the common signals measured by the reference sensors and the sensor under test that can arise from factors such as their different locations and their different responses. The Multi-Channel Coherent Subtraction script searches for the optimal $T_{j}$ for each additional sensor used by taking into account the coherence between all the sensors\cite{Allen1999}. Afterwards, the Multi-Channel Coherent Subtraction script applies a statistical coherence correction factor to account for the statistical chance that random signals (e.g. noise) are coherent\cite{Allen1999,Bendat2000,Hua2005}.
\section{Results}
\label{Results}
\subsection{Set-up}
\label{Set-up}
\begin{figure}
	\centering
	\includegraphics[trim = 0cm 0cm 0cm 0cm, clip, width=8.5cm]{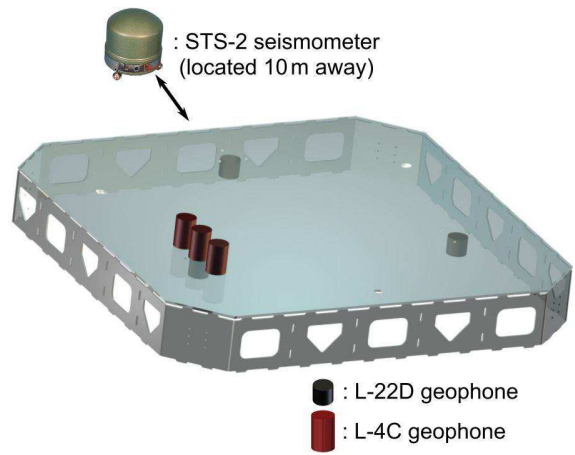}
	\caption{The sensor arrangement for the huddle tests. Three pre-installed L-22Ds that are arranged in a triangular, the STS-2 seismometer standing 10\,m apart and three additionally installed L-4Cs were used. Except of the STS-2 seismometer the sensors were all isolated from ground motion by the AEI-SAS.}
	\label{Huddletest_setup}
\end{figure}
As the goal was to measure the sensor noise, the geophones were placed in a seismically quiet location. In this test we used the AEI-SAS\cite{Wanner2013,Wanner2012,Bergmann2017} which passively isolates an optical table from ground motion with additional active damping of its main resonances. Nevertheless the sensor signals were still dominated by residual motion which is why the huddle tests were performed.\\\\
The geophone arrangement is shown in FIG.~\ref{Huddletest_setup}. The AEI-SAS optical table already had three L-22Ds pre-installed inside the table in a triangular arrangement. Three additional L-4Cs were installed above of the existing L-22Ds onto the table top. An STS-2 seismometer was also used as an additional sensor, which was located on the ground approximately 10\,m away. The sensor under test is one of the L-4Cs for the L-4C huddle test and it is the L-22D placed right below the L-4Cs for the L-22D huddle test.\\\\
Data was acquired using 16-bit analog-to-digital converters (ADCs) with a sampling rate of 1\,kHz (downsampled from 64\,kHz) and using anti-aliasing filters, consisting of two Sallen-Key filters with corner frequencies of 10\,kHz. Time series of 3000\,s were recorded simultaneously for each sensor in the huddle test. 

\subsection{Seismically quiet location}
\label{Seismically quiet location}
\begin{figure}
	\centering
	\includegraphics[trim = 0cm 0cm 0cm 0cm, clip, width=8.5cm]{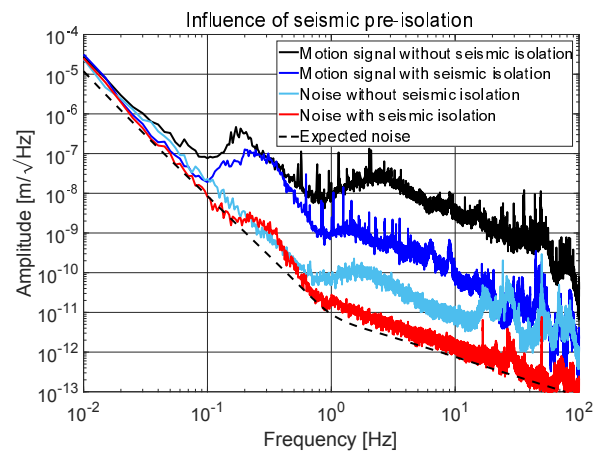}
	\caption{The influence of the seismic isolation for the huddle test results with the raw signals (motion and noise) of the target sensor shown for comparison. The measured noise without seismic isolation contains various contributions of ground motion at frequencies above $\approx$\,1\,Hz that could not be subtracted. The 1/$f^2$ slope above the main resonance ($\approx$\,0.3\,Hz) of the passive seismic isolation and the active damping of this resonance as well as the microseismic peak ($\approx$\,0.2\,Hz) is visible.}
	\label{seismic_isolation}
\end{figure}
To investigate the influence of the seismic isolation for huddle test results, a huddle test was performed with the sensors placed on a regular optical table that is rigidly connected to ground. FIG. \ref{seismic_isolation} shows the seismic motion on top of the isolated and the non-isolated optical table as well as the measured sensor noise on both. It can be seen that the noise measurement without seismic isolation still contains substantial contributions from ground motion above $\approx$\,1\,Hz. In contrast, when the sensors were placed on the AEI-SAS, the measured noise is lower and contains fewer peaks. In theory, the seismic isolation should not make a difference for the huddle test since the ground motion is detected coherently and subtracted. However, in reality the ground motion is not detected with perfect coherence by the different sensors. In particular, horizontal motion acting on the geophones gets converted into a vertical signal due to some shaking of the suspended mass. This might be detected incoherently as small mechanical differences of the geophone interior can influence this conversion. 
\subsection{Multiple sensors}
\label{Multiple sensors}
\begin{figure}
	\centering
	\includegraphics[trim = 0cm 0cm 0cm 0cm, clip, width=8.5cm]{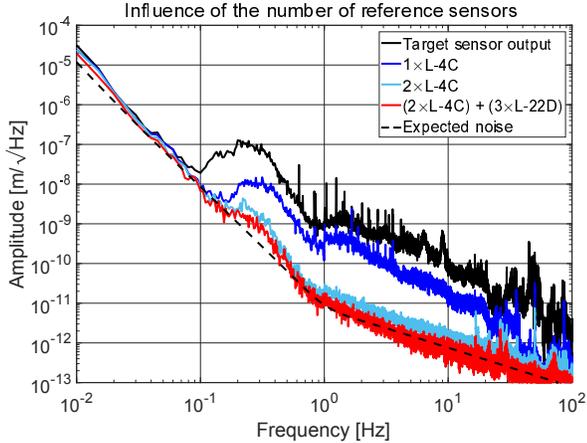}
	\caption{The influence of the number of reference sensors for a huddle test. The more reference sensors are used, the more coherent signals can be subtracted from the target signal.}
	\label{numer_reference_sensors}
\end{figure}
The optimal number of sensors used for the Mutli-Channel Coherent Subtraction was also investigated. FIG. \ref{numer_reference_sensors} shows the output of the target L-4C sensor and the residual output as the coherent information from additional reference sensors is subtracted. These measurements were done on the AEI-SAS. It can be seen that as more reference sensors are used, the lower the residual output from the target sensor becomes. When only a single reference sensor is used, the sensor noise measurement still contains significant contributions from actual ground motion. It should be noted that the peak at $\approx$\,0.3\,Hz is still visible in the measurement even with the largest number of reference sensors. This is exactly the resonance frequency of the vertical isolation stage of the AEI-SAS. Although this resonance was actively damped, this still results in a vertical motion that is larger than the ground motion. In addition, the horizontal motion of the AEI-SAS is large at this frequency due to cross-coupling between vertical and horizontal degrees of freedom. As explained above, horizontal motion might be detected incoherently and thus can not be subtracted from the target sensor. The use of additional horizontal reference sensors might have improved the huddle test accuracy in this frequency regime further.

\subsection{Comparison between L-22D and L-4C}
\label{Comparison between L-22D and L-4C}
\begin{figure}
	\centering
	\includegraphics[trim = 0cm 0cm 0cm 0cm, clip, width=8.5cm]{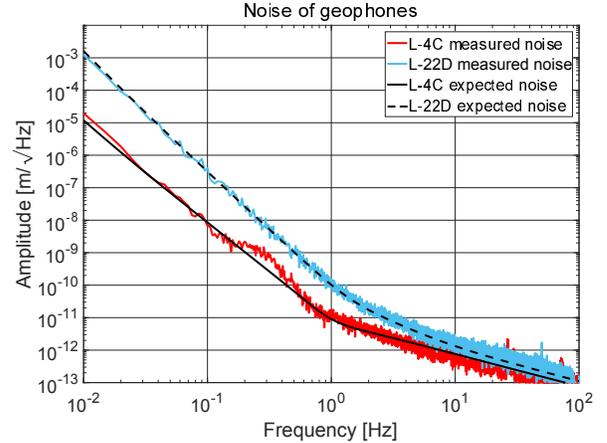}
	\caption{Comparison of the measured L-4C and L-22D noise and the theoretically calculated noise.}
	\label{NoiseMeasurement}
\end{figure}

FIG. \ref{NoiseMeasurement} shows the best results of the huddle tests for the two geophones and their amplifier circuits using the AEI-SAS and all available reference sensors. The measured L-22D and L-4C noise curves overlap very well with the predictions over the entire bandwidth of interest. The L-4C noise is very close to the prediction except for frequencies between $\approx$\,0.2\,Hz and 0.6\,Hz. For the reasons explained above this deviation is assumed to be due to an incoherent cross-coupling of resonantly enhanced ground motion. From the noise budget (FIG. \ref{L4Cnew}), it can be seen that the two dominant noise sources are the Johnson noise and the operational amplifier voltage noise. While the voltage noise might be further reduced by finding a lower noise amplifier, the Johnson noise is a fundamental limit of the geophone. Consequently, an improvement of more than a factor of $\approx$\,$\sqrt{2}$ is not possible for the L-4C geophones.\\\\
The overall noise of an L-4C and the amplifier electronics described in the supplementary material (section \ref{Supplementary material}) reaches down from low frequency (0.01\,Hz) with a slope of $1/f^3$ to a value of $\approx$\,$10^{-11}\,\text{m}/\sqrt{\text{Hz}}$ at 1\,Hz, and falling off at higher frequencies with a slope of $1/f$. The L-4C geophones and their amplifier electronics will be used in the future for the active control of the AEI-SAS improving the active isolation performance.

\section{Conclusion}
\label{Conclusion}
In this paper it is shown that the measurement of sensor noise of seismic sensors by performing a huddle test can be significantly improved by performing the tests in a seismically quiet environment and by using larger numbers of reference sensors to remove the remaining background signal. Utilizing these two techniques the sensor noise of two geophone-amplifier configurations were measured down to a frequency of 0.01\,Hz. These measurements also show that the combination of an L-4C geophone and low noise operational amplifiers OPA188 can result in a sensor which was characterized to be near Johnson noise limited. The measured sensor noise of this configuration reaches a value of  $10^{-11}\,\text{m}/\sqrt{\text{Hz}}$ at 1\,Hz. Both noise measurements agree remarkably well with their calculated noise budgets.

\section{Supplementary material}
\label{Supplementary material}
Supplementary material 1 shows the schematics of the L-22D and the L-4C preamplifier.\\ Supplementary material 2 shows the MATLAB code of the Multi-Channel Coherent Subtraction script.\\\\

\textbf{\LARGE{Acknowledgments}}\\\\
The authors gratefully thank the International Max Planck Research School
(IMPRS) on Gravitational Wave Astronomy and QUEST, the Center for Quantum
Engineering and Space-Time Research for their support. This project has received funding from the European Union’s Horizon 2020 Research and Innovation Programme under the Marie Sklodowska-Curie grant agreement Number 701264. We also thank Brian Lantz for his very helpful comments on this paper.\\

\nocite{*}
\bibliography{literature}

\end{document}